\documentclass[aps,prl,twocolumn,groupedaddress,amsmath,amssymb]{revtex4}
\usepackage[T1]{fontenc}
\makeatletter

\newcommand{\noun}[1]{\textsc{#1}}
\newcommand{\Unit}{\openone}
\newcommand{\dirac}[1]{\vert #1 \rangle }
\newcommand{\carid}[1]{\langle #1 \vert }

\makeatother

\begin{document}

\preprint{EIQU preprint}

\title{There is no simulation of n-qubit operations by a single Hamiltonian
with 2-spin interaction}

\author{Torsten Asselmeyer-Maluga, Matthias Kolbe, and Helge Rosé}
\affiliation{FhG FIRST, Kekuléstr.~7, 12489 Berlin, Germany}
\date{\today}

\begin{abstract}
Today's devices for quantum computing are still far from implementing useful and powerful quantum algorithms. Decoherence and the wish to resist the effects of errors in a system of quantum bits incurs a lot of overhead in the number of gates and qubits. From a theoretical perspective, controlled quantum simulation raises the hope to simulate the unitary quantum operationes generated by a Hamiltonian with 3-body interaction with a suitably designed element that is constructed of only 2-body interactions. That replacement would happen without any additional gates, and its possibility would be due to the ambiguity of the unit element of the Lie group connected with the algebra of traceless hermitian matrices. We show that this hope is void, and give a general proof for this for any order of interaction.
\end{abstract}

\maketitle

\section{introduction}

The outstanding properties of exponentiality, fast access, universality, branching and interference in connection with complex amplitudes as a whole give reason to hope that new classes of problems can be solved with quantum computation in contrast to the conventional von-Neumann machines \cite{Deu92,Ber97}. However, in all those optimistic visisons of straight-forward developments of quantum computing, one has to face the unavoidable effects of decoherence. Nowadays, in the presence of the broadly investigated field of error correction, the view in dealing with the difficulties of decoherence could be too optimistic. 
In particular, a high amount of entanglement, which seems to be essential for good quantum algorithms, is very sensitive for decoherence effects \cite{Pres98b}. Of course, it is indeed possible to protect quantum states against unwanted influences with error correcting codes and fault-tolerant quantum algorithms, but there is a high price to be paid in an enormous overhead of gates and qubits required to store and proceed redundant information.
This overhead is acceptable however as long as the error rate per gate, the accuracy threshhold, is under a certain critical value \cite{Kni96,Pres97}. 

A rough estimation of the gate number required for implementing the quantum fourier transformation (see \cite{NS} for instance), which is a major component in Shor's well known factoring algorithm \cite{Shor97}, shows that for $n$ qubits no fewer than $n$ Hardarmard gates  plus $\frac{1}{2}(n^2-n)$ controlled phase gates are required \cite{NS}. For the more essential Modulo operation, many more gates, at the order of $~n^3$ \cite{vanMet05}, have to be considered. So for a useful factorization of a big number, say of 1000 bits, it is very likely that hundreds of millions of gates need to be applied, and this number will be further increased massively (albeit polynominally) by error correction. Very recent investigation shows how architecture \cite{vanMet05} and error correction \cite{Dev04} affect the computation time of Shor's algorithm. Ion-trap experiments, being good realization candidates, show gate operation times from $10^{-14}\,\mathrm{s}$ up to microseconds, whereas typical decoherence time $\tau$ is about one second. Recent experiments by the group of Rainer Blatt in Innsbruck \cite{Kre04} come up with $\tau=10\,\mathrm{s}$ for ${}^{40}$Ca-Ions. In nuclear magnetic resonance (NMR) experiments, another very good candidate, decoherence time could be even much longer, up to $10^8$ seconds, but the operation time per gate also increases to milliseconds \cite{NS}. Lloyd mentions decoherence time for NMR in such a long range as years \cite{Lloyd99} under optimal conditions. Of course, these values show that up to $10^{14}$ operations and more might be possible in these systems, but only under really optimal circumstances. However, the mutually influences of decoherence, error correction, gate count, qubit count, accuracy threshhold and operation time (all in strong architectural dependency) results in a situtation that seems to be far from implementing really practical and powerful quantum algorithms within large quantum systems and realistic decoherence times.

But not only these rather technical restriction have to be taken into account for realizing a quantum computer. As we shall outline, the operations that act on multiple qubits simultaniously, which occur in various algorithms, are generated from unphysical Hamiltonians as well. The standard construction method of taking logarithms of unitary operators reveals Hamiltonians with multipartite interaction terms. Of course, arbitrary $n$-bit quantum gates can be expressed as compositions of 2-bit quantum gates, and this universality is well developed in its basics \cite{Bar95c,Deu95,diVin94,Sle95,Bar95b,diVin95a,Lloyd95} and optimizations (see for instance the work of Zhang et al.\ \cite{Zha03a,Zha03b,Zha04,Zha05}). Unfortunately, this happens at the cost of increasing gate numbers, which is critical as outlined above. Wouldn't it be nice to have a replacement of such Hamiltonians by a single realistic one? There appears to be hope to do so, as we will discuss in the next section, but in the rest of the paper we shall prove the contrary.

\section{Constructing Hamiltonians from unitary operations}
\label{hamiltonconstruction}

In the following we will introduce the problem of the representation
of a unitary operation (as an element of the unitary group $U(2^n)$)
by the exponential of a suitable Hamiltonian (as an element of the Lie
algebra $\mathfrak{u}(2^n)$). As an example we consider the
CNOT operation represented by the unitary matrix
\[
\left(\begin{array}{cccc}
 1 & 0 & 0 & 0\\
 0 & 1 & 0 & 0\\
 0 & 0 & 0 & 1\\
 0 & 0 & 1 & 0\end{array}\right)\]
also given by the Lie algebra element
\[
it\,H=i\frac{\pi}{4}(\Unit-\sigma_{z})\otimes(\Unit-\sigma_{x})
\]
 where $\sigma_{x}$, $\sigma_{z}$ are the Pauli matrices and $\Unit$
is the $2\times~\!\!2$ unit matrix. From the physical point of view, every
unitary operation must be generated by a Hamiltonian with respect
to a special time. In our example we have
\[
CNOT=\exp(iHt)
\]
 with $t=\pi/4$ and $H_{CNOT}=(\Unit-\sigma_{z})\otimes(\Unit-\sigma_{x})=\sigma_{z}\otimes\sigma_{x}-\sigma_{z}\otimes\Unit-\Unit\otimes\sigma_{x}+\Unit\otimes\Unit$. 
This Hamiltonian can be physically interpreted as a spin system with 2-spin interaction
given by $\sigma_{z}\otimes\sigma_{x}$ in an exterior field $\sigma_{z}\otimes\Unit+\Unit\otimes\sigma_{x}$.
For a 3-qubit operation like the Toffoli gate we obtain ($t=\pi/8$):
\begin{align*}
H_T&=(\Unit-\sigma_{z})\otimes(\Unit-\sigma_{z})\otimes(\Unit-\sigma_{x})\\
&=\Unit\otimes(\Unit-\sigma_z)\otimes(\Unit-\sigma_x)\\
&\qquad-(\sigma_z\otimes\Unit\otimes\Unit) + (\sigma_z\otimes\Unit\otimes\sigma_x)\\
&\qquad+(\sigma_z\otimes\sigma_z\otimes\Unit) - (\sigma_z\otimes\sigma_z\otimes\sigma_x)
\end{align*}
and thus a 3-spin interaction $\sigma_{z}\otimes\sigma_{z}\otimes\sigma_{x}$.
But an interaction between 3 constituents is artificial in nature 
\footnote{It follows from first principles that matter is represented by fermions
given mathematically as Dirac spinors and fulfilling the Dirac equation.
Interaction between two fermions is only given by introducing a gauge
field which coupled to the Dirac spinor. But that kind of coupling
leads to a 2-particle interaction.} and only possible under very restricted conditions. 

Representations of Hamiltonians in terms of eigenenergies may be
related to representations in terms of Pauli spin matrices $\sigma_z$
much more generally \cite{Tse99}.  Assume an Hamiltonian in its
eigenstructure. It can be written as
\[
H = \sum_{k=0}^{\infty} \varepsilon_k\dirac{\psi_k}\carid{\psi_k}.
\]
Here, the $\dirac{\psi_k}$ are the complete set of orthogonal
eigenstates and the $\{\varepsilon_i\}$ are the energy eigenvalues.
For simu\-lation in an n-qubit system let us truncate the sum to the
first $2^n$ ernergie levels. Then we have
\begin{align*}
H &= \sum_{k=0}^{2^n-1} \varepsilon_k\dirac{\psi_k}\carid{\psi_k}\\
&= \sum_{k=0}^{2^n-1}
\alpha_k\left(\sigma_z\right)^{\nu_1^k}\otimes\left(\sigma_z\right)^{\nu_2^k}\otimes\dotsb\otimes\left(\sigma_z\right)^{\nu_n^k},
\end{align*}
where the $\{\alpha_i\}$ are real numbers representing coupling
strength and the $\{\nu_i^k\}$ are the binary representation digits
for the integer k, thus take on the values $\{0,1\}$. It turns out
that the vectors $\boldsymbol{\varepsilon}$ and $\boldsymbol{\alpha}$
are related by the matrix equation $\boldsymbol{\varepsilon} =
\boldsymbol{M}\boldsymbol{\alpha}$ with $\boldsymbol{M}$ as the
Hardamard matrix for n qubits. For example, for two qubits, we have:
\[
\boldsymbol{M}=
\left(\begin{array}{rrrr}
1&1&1&1\\
1&-1&1&-1\\
1&1&-1&-1\\
1&-1&-1&1
\end{array}\right).
\]
Once the arbitrary Hamiltonian is expressed in terms of many-body
interactions $\sigma_z\otimes\sigma_z\otimes\dotsb\otimes\sigma_z$, it
can be broken down in terms of available external and internal
(two-body) Hamiltonians. Control theory enables us, furthermore,
to simulate arbitrary Hamiltonians by those that are predetermined 
by some appropriate experiment \cite{Woc01}.

At this point we want to briefly introduce the decomposition
techniques of Khaneja et~al.\ \cite{Kha00, Kha01} as an interesting
theoretical approach to universality. Assume an element of the
special unitary group describing qubit evolution, $U\in SU(2^n)$. It
is always possible to decompose it into $U=K_1AK_2$ where $K_1,K_2 \in
SU(2^{n-1})\otimes SU(2^{n-1})\otimes U(1)$ as long as A is an element
of the so-called Cartan subalgebra of the Riemann symmetric space
\[
\frac{SU(2^n)}{SU(2^{n-1})\otimes SU(2^{n-1})\otimes U(1)}\,.
\]
One should note that this is recursive, because then we can further
decompose $K_1$ and $K_2$ in $SU(2^{n-2})\otimes SU(2^{n-2})\otimes
U(1)$ and so on, down to elements of $SU(2)\otimes SU(2)$. This
decomposition ist based on the parametrization of $SU(2^n)$ with
canonical parameters of the second kind \cite{Sagle}. Suppose $U \in
SU(2)$, then we can express any element in two ways
\begin{samepage}
\begin{enumerate}
\item $U=\exp\bigl[-i(\alpha_1\sigma_x + \alpha_2\sigma_y + \alpha_3\sigma_z)\bigr]$
\item $U=\exp(-i\beta_1\sigma_x)\exp(-i\beta_2\sigma_y)\exp(-i\beta_3\sigma_z)$
\end{enumerate} 
\end{samepage}
with $\alpha_i,\beta_i\in\mathbb{R}$. This coincides with the two kinds of
canonical parameters, and it would be promising to look for an
decomposition technique that suits the first of the above standart
parameterizations. Actually, such a decomposition is the goal of our
investigation to implement a 3-qubit operation with a single
2-particle Hamiltonian. Remember that in the case of $SU(8)$, terms of
3-particle interations (like in the Toffoli-Hamiltonian above) and
terms of 2-particle interactions are orthogonal in their algebra
$\mathfrak{su}(8)$ as they are different basis elements. Based on that
fact our goal may appear out of reach, but there is an ambiguity in
the exponential!

Consider the eigenvalues $\lambda_{1},\ldots,\lambda_{2^{n}}$ of the
Hamiltonian $H$ and the unitary matrix $A$ of eigenvectors
diagonalizing $H=A(diag(\lambda_{1},\ldots,\lambda_{2^{n}}))A^{+}.$
Then the exponential $\exp(iHt)$ can be written as\[
\exp(iHt)=A(diag(e^{i\lambda_{1}t},\ldots,e^{i\lambda_{2^{n}}t}))A^{+}\]
and we can shift every eigenvalue $\lambda_{k}t+2\pi n_{k}$ by an
integer $n_{k}$ so that the exponential is unchanged. We denote this
shift by
\[N=2\pi A(diag(n_{1},\ldots,n_{2^{n}}))A^{+}\]
with $[H,N]=0$ and $\exp(iN)=\Unit$. Thus, we obtain
\[
\exp(iHt)=\exp(iHt+iN)\,.
\]
Now we consider a 3-qubit system. Let $H$ be a Hamiltonian with
3-spin interactions and $h$ a Hamiltonian with 2-spin interactions
defined for a 3-qubit system. By the ambiguity above, there is perhaps
a shift $N$ so that 
\begin{align*}
Ht=ht'+N&\Longrightarrow\exp(iHt-iht')=\exp(iN)=\Unit\\
&\Longrightarrow\exp(iHt)=\exp(iht')=U
\end{align*}
and we ask for the existence of such a shift $N$ with
$[(Ht-~\!\!ht'),~N]=0$.  
We want to emphasize here that this is not a trivial question
and it is not obvious what's coming out at the end.  Anyhow, we must
dispel the hope that such an $N$ exists and we will prove it for any
order in the next section.

\section{The No-Go Theorem}

In this section we will consider the following situation: an $n$-qubit
system with state space $\mathbb{C}^{2^{n}}$ and a unitary operation
$U$ lying in $U(2^{n})$. Furthermore, we have a Hamiltonian $H$
with $n$-spin interaction and a Hamiltonian $h$ with $(n{-}1)$-spin
interaction. Assume
\[
U=\exp(iH)\,,
\]
 then we will show that there is no Hamiltonian $h$ with
\[
U=\exp(iH)=\exp(ih)\,,
\]
i.e. every unitary operation $U\in U(2^{n})$ can only be represented
by a Hamiltonian with $n$-spin interaction. For a warm-up example, we
start with the first non-trivial case $n=2$ and state that no unitary
2-qubit operation $U$ generated by $H$ can be also generated by $h$.
To prove this we begin with the assumption that by definition the
exponential of a 2-spin interaction given by
$H=\sigma_{i}\otimes\sigma_{j}$, $i,j\in\{x,y,z\}$, can never be
decomposed as
\begin{equation}
\exp(\sigma_{i}\otimes\sigma_{j})\not=A\otimes B\qquad A,B\in U(2)\,,\label{assume-n-2}
\end{equation}
otherwise the 2-qubit operation is decomposible by 1-qubit operations.
Furthermore, every 1-spin ``interaction'' is given by
$h=\sigma_{j}\otimes\Unit+\Unit\otimes\sigma_{i}$, and we have
\begin{eqnarray*}
\exp(ih) & = & \exp\bigl[i(\sigma_{j}\otimes\Unit+\Unit\otimes\sigma_{i})\bigl]\\
 & = & \exp(i\sigma_{j}\otimes\Unit)\exp(i\Unit\otimes\sigma_{i})\\
 & = & \exp(i\sigma_{j})\otimes\Unit)(\Unit\otimes\exp(i\sigma_{i})\\
 & = & \exp(i\sigma_{j})\otimes\exp(i\sigma_{i})\,,
\end{eqnarray*}
 but that contradicts (\ref{assume-n-2}). 

 We showed that no non-trivial unitary operation generated by 2-spin interactions
 can also be generated by 1-spin interactions (which physically represents
 an exterior field that acts on the spin system). Ok, this might be
 no suprise because no interaction whatsoever was allowed here.

\medskip 
\textbf{No-Go-Theorem for Hamiltonian representations}:
\textit{No unitary $n$-qubit operation $U$ ($n>2)$ generated by a Hamiltonian
$H$ with $n$-spin interactions can be generated by a Hamiltonian
$h$ with $(n{-}1)$-spin interactions as well so that the relation
\begin{equation}
U=\exp(iH)=\exp(ih)\label{assumption}
\end{equation} 
is fulfilled.
}

\noun{Proof:} Consider a Hamiltonian
$H_{2^{n}}$ as an element of the Lie algebra $\mathfrak{u}(2^{n})$ and
a Hamiltonian $H_{2^{n}-1}$ as an element of the Lie algebra $\mathfrak{u}(2^{n}-1)$.
We will prove that an element $U$ of the unitary group $U(2^{n})$
generated by $H_{2^{n}}$ can never be generated by $E(H_{2^{n}-1})$
with respect to all embeddings $E:\mathfrak{u}(2^{n}-1)\to\mathfrak{u}(2^{n})$.
Then the theorem follows by using $2^{n-1}$ times that result.
In the following we use the abbreviation $k=2^{n}.$

Consider a family of Hamiltonians $H_{k}(a_{1},\ldots,a_{k^{2}})$
parametrized by $k^{2}$ parameters which is the dimension of the
Lie algebra $\mathfrak{u}(k)$, i.e. we have a map $H_{k}:U(k)\to\mathfrak{u}(k)$
from the coordinates of the Lie group (seen as smooth manifold with
group operation) to the Lie algebra (seen as tangent space of the
Lie group) \cite{Sagle}. The tangent bundle $TU(k)$ of
the Lie group is trivial, i.e. $TU(k)=U(k)\times\mathfrak{u}(k)$.
Thus the map $H_{k}$ extends to a map $H_{k}:U(k)\to TU(k)$, i.e.
$H_{k}$ is a vector field on $U(k)$. By the same argument we
can interpret $H_{k-1}$ as a vector field on $U(k-1)$. By the simple
algebraic argument of linear independence, both vector fields $H_{k}$
and $E(H_{k-1})$ disagree, i.e.
\begin{equation}
H_{k}\not=E(H_{k-1})\,.\label{assumption-lin-indi}
\end{equation}
By using the assumption (\ref{assumption}) and the linear independence
(\ref{assumption-lin-indi}) we have
\[
\exp(iH_{k})=\exp\bigl[iE(H_{k-1})\bigr]\Longrightarrow H_{k}-E(H_{k-1})=N_{k}\,,
\]
where $N_{k}$ is a vector field that depends on $k^{2}-(k-1)^{2}=2k-1$
parameters. 
Then we can interpret the vector field $N_{k}$ as vector field on
$U(k)$ modulo $U(k-1)$ or as vector field on the coset space
$U(k)/U(k-1)$, i.e. for a fixed $H_{k}$, the variation of $N_{k}$ with
respect to $E(H_{k-1})$ is expressed by this coset space
$U(k)/U(k-1)$. The definition of this space is given by the fact that
the group $U(k-1)$ acts on $U(k)$, and two elements $g_{1},g_{2}\in U(k)$ 
are said to be equivalent if and only if an element $G\in
U(k-1)$ with $g_{2}=Gg_{1}$ exists. Then the equivalence classes are
denoted by $U(k)/U(k-1)$. It is a well-known fact \cite{Nakahara} that
$U(k)/U(k-1)=S^{2k-1}$, i.e.\ the $(2k-\nobreak 1)$-dimensional sphere. Now, if
we can show that the vector field $N_{k}$ vanish at some point then we
can shift this vanishing point at every place to show that
\[
H_{k}-E(H_{k-1})=0\,,
\] 
thus contradicting the linear independence of $H_{k}$ and $E(H_{k-1})$ (see
(\ref{assumption-lin-indi}) above). Thus we are looking for the
existence of a non-vanishing vector field on $U(k)/U(k-1)=S^{2k-1}$
that represents $N_{k}$. By a famous mathematical result of Adams
\cite{Adam62}, there is only a non-vanishing vector field on
$S^{2k-1}$ for $k=1,2,4$.  The vector fields on all other spheres
vanish in one point, which would contradict (\ref{assumption-lin-indi}). 
The cases $k=1,2$ are trivial, and $k=4$ is covered by our warm-up example. 
That completes the proof.
\textbf{qed}

\section{Conclusion}

Based on the fact that there is an ambiguity of the unit element of a
Lie group connected with its Lie algebra, the hope is raised that
elements of $SU(2^n)$ generated by Hamiltonians of
$\mathfrak{su}(2^n)$ carrying $n$-body  interactions can
also be generated by Hamiltonians carrying at most $(n{-}1)$-body
interactions. The ambiguity can be interpreted as equivalence classes
represented by multidimensional spheres. Therefore, by transferring this 
problem to a geometrical view and treating Hamiltonians as vector
fields on the group, we could show that the hope of replacing unphysical
multi-particle interactions is void. The central idea of the proof is the 
theorem of Adams \cite{Adam62} about vanishing vector fields on spheres.
The degree of interaction, which is produced by the logarithm of a unitary operation, 
cannot be reduced. Thus, for $n$-qubit operations, $n$-body interactions are needed. 
The only way to avoid those interactions is the decomposition of unitary operations in terms
of universal 2-qubit gates for the price of a higher number of operations.
With this insight we want to challenge the theory of adiabatic quantum
computing \cite{Aha04}, which heavily relies on the implementation of
3-body interactions. Therefore, we want to notice that the realization
of the ideas of adiabatic computing is daring if not impossible.

\begin{acknowledgments}
  This work was supported by the BMBF under the project number
  01IBB01A.  We thank Dominik Janzing, Andreas Schramm and Martin
  Wilkens for useful hints and discussions.
\end{acknowledgments}

%\bibliography{./quantum}

\end{document}